Explanation for Cancer in Rats, Mice and Humans due to Cell Phone Radiofrequency Radiation

Bernard J. Feldman

Department of Physics and Astronomy

University of Missouri-St. Louis

Abstract

Very recently, the National Toxicology Program reported a correlation between exposure to whole body 900 MHz radiofrequency radiation and cancer in the brains and hearts of Sprague Dawley male rats. This paper proposes the following explanation for these results. The neurons around the rat's brain and heart form closed electrical circuits and, following Faraday's Law, 900 MHz radiofrequency radiation induces 900 MHz electrical currents in these neural circuits. In turn, these 900 MHz currents in the neural circuits generate sufficient localized heat in the neural cell axons to shift the equilibrium concentration of carcinogenic radicals to higher levels and thus, to higher incidences of cancer. This model is then applied to mice and humans.

Very recently, the National Toxicology Program (NTP) reported a correlation between exposure to whole body radiofrequency radiation and cancer in Sprague Dawley male rats.[1] The experiment consisted of irradiating Sprague Dawley rats and B6C3F1 mice with 900 MHz radiation with four different intensities: 0 W/kg, 1.5 W/kg, 3 W/kg and 6 W/kg. The frequency of 900 MHz was chosen because it is typical for use in cell phones and other wireless devices. The exposure times were 10 minutes on and 10 minutes off for 18 hours a day, resulting in a total exposure of nine hours daily. The animals were exposed daily from in utero until two years of age. The animals were monitored so that exposure was at a low non-thermal or non-heating level. Groups of 90 animals were used for each species, sex and intensity.

They reported the following results: 1. Excess cancers were found only in male rats but not in female rats, male mice or female mice. 2. Only brain cancers (gliomas and brain lesions) and heart cancers (schwannomas) were found. Schwannomas are cancers of the neural cell sheaths. 3. The incidence of cancer in male rats increased as the 900 MHz radiofrequency intensity increased from 0 to 6 Watts/kilogram. In particular, no cancers were found in any animal that was not exposed to radiofrequency radiation (0 W/kg). 4. Even at the highest radiofrequency power, 6 W/kg, this power was insufficient to significantly increase the rats' average body temperature by more than one degree Centigrade. 5. The rats exposed to radiofrequency radiation lived longer than those rats that were not exposed. 6. Analysis of bioassays showed that "male rats are more sensitive to chemical carcinogenesis compared to female rats." 7. A small minority of the reviewers of this study questioned the statistical significance of these results. 8. One of the collaborators in the NTP also mentioned that some previous studies had found similar brain and heart excess cancers in humans due to radiofrequency radiation[2].

Theoretical understanding of the interaction between animals and electromagnetic radiation has a long and complicated history. Physicists in general have been very skeptical of any connection between any non-ionizing radiation and cancer. The classic paper by R. Adair on weak extremely low frequency (60 Hz) electromagnetic fields concluded that "such interactions are too weak to have a significant effect on human biology at the cell level."[3] Adair applied Faraday's Law to a single cell radiated with weak 60 Hz electromagnetic fields and concluded that the induced electric field is small compared to thermal noise induced electric fields. Even this author expressed skepticism about cell phone radiation (900 MHz) causing cancer by using an analogy with Einstein's theory of the photoelectric effect—the 900 MHz photon energies are about a million times less than the energy needed to break chemical bonds.[4] D. Eichler applied Faraday's Law to a closed neural circuit and calculated large induced electric fields across the synapse membrane.[5] Recently, Barnes and Greenebaum proposed that weak static and high frequency magnetic fields could change the recombination rate of radical pairs and thus change the concentration of carcinogenic radicals like $O_2^-$ in cells.[6] Panagopoulos, Johansson and Carlo suggested that high frequency electric fields exert electrostatic forces on the cell membrane, disrupting the functioning of the ion channels.[7]

The following explanation for the NTP results is proposed. The neurons around the brain and heart form closed electrical circuits and, following Faraday's Law, 900 MHz radiofrequency radiation induces 900 MHz electrical currents in these neural circuits. Given that the axons of these neural cells are about one micron thick, these 900 MHz currents in the neural circuits could generate sufficient localized heat in the axons of the neural cells to significantly raise the temperature of the neural and neighboring cells and shift the equilibrium concentration of carcinogenic radicals in these cells to higher levels and thus, to higher incidences of cancer.

Consider a neural circuit on the surface of the brain or the heart in the shape of a circle of radius r. From Faraday's Law, the induced voltage, V, in the neural circuit is equal to minus the time (t) derivative of the magnetic flux crossing the closed circuit. Assuming that the radio frequency magnetic field is $B = B_o \cos\omega t$ where $B_o$ is a constant and $\omega/2\pi$ = 900 MHz, then V = $\omega B_o \sin\omega t \, \pi r^2$ and is proportional to $B_o$, $r^2$ and $\omega$. The resistance of the neural circuit, R, is proportional to its circumference and thus proportional to r, since the width of neural cell axons is largely independent of the size or type of animal. Then the heat generated in the neural circuit, P = $V^2/R$, is proportional to $B_o^2$, $\omega^2$, and $r^3$ and the heat generated per unit length of the neural circuit is proportional to $B_o^2$, $\omega^2$ and $r^2$.

Now the question arises: how could this excess localized heat cause cancer? Again, let us consider a simple model consisting of carcinogenic radicals, like oxidants, O, and antioxidants that scavenge these radicals, A. Inside cells, chemical reactions occur that convert food into useful chemicals, heat, and muscular motion. Carcinogenic radicals are created as waste products of these chemical reactions. In order for the animal to protect itself from these carcinogenic radicals, the cells create antioxidants. The antioxidants bind to the carcinogenic radicals forming harmless molecules, OA, which then diffuse to nearby veins and are removed from the body via the kidney and urinary system. The concentrations of O and A, [o] and [a], are in approximate equilibrium with the concentration of OA, [oa]. In other words, the cell's rate of production of O equals the cell's rate of production of A, which in turn equals the rate of removal of OA from the cell. One can write down the following chemical reaction.

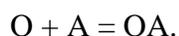

O + A = OA.

From the law of mass action, at equilibrium,

[oa]/[o][a] = K(T)

and

K(T) = C exp(-G/RT)

where K is the equilibrium constant for this chemical reaction, T is the temperature in Kelvin, C is a constant, G is the Gibbs free energy of the chemical reaction, and R is the gas constant. G is negative for exothermic reactions. Notice as the temperature goes up, K(T) goes down, and [o] and [a] go up. An increase in [o] is believed to be connected with an increase in the cancer rate.

An alternative possibility is that the increased temperature denatures antioxidants (loses their structure and thus their ability to function), leaving higher concentrations of carcinogenic radicals. In humans, proteins start to denature when the body temperature is above 42º C, not much higher than the body temperature of 37º C.

This model is consistent with the NTP experimental observations. The cancers are found near neurons, found in large organs that are surrounded by or largely consist of neural cells, like the brain and the heart, but not found in small organs like the thyroid, prostate or the kidney or in large organs not surrounded by or consisting of neural cells like the pancreas, intestine, lung or the liver. The incidence of cancer increased with the radiofrequency intensity which is proportional to $B_o^2$. Even though the rats' average body temperature does not increase significantly, there could be significant local heating of these neural circuits undetected by the NPT. Given that male Sprague Dawley rats are on average about 60% heavier (and thus about 60% larger) than female Sprague Dawley rats at 15 weeks of age, the model does predict a higher incidence of cancer in male rats than female rats, but sex hormones play a much bigger role in this difference between male and female rats' cancer rates.

This model is also consistent with the observation that no cancers were found in mice. Since B6C3F1 mice are about a factor of three smaller in size than the Sprague Dawley rats, our theory predicts that the incidence of cancer in mice should be reduced by a factor of nine compared to rats. Given that the incidence of cancer in male rats was barely statistically significant, a decrease of a factor of nine would predict no cancers observed in male mice. This observation of no cancer in male mice strongly argues against any cancer mechanism that is at the cellular or molecular level, like the ones in references [6] and [7]. Given that nerve cells are about the same size in mice and rats and the genetics and biology of mice and rats are extremely similar, a cellular or molecular theory of cancer would predict cancer in male mice, if there is cancer in male rats.

Finally, the NTP did comment on the observation that irradiated rats lived longer than non-irradiated rats. They suggested that it is related to the observation that calorie–restricted animals live longer on average. This is also consistent with the above model since an animal that is heated will consume less food, produce less carcinogenic radicals and thus live longer.

There are two other experimental results that support of this model. C.-K. Chou, et al. measured the specific absorption rates of 2,450 MHz radiation in different parts of rats' brains.[8] The experiment consisted of using two dead rats, one exposed to radiation and the other not, whose small portions of brain tissue were removed right after radiation and placed in a calorimeter to measure the specific absorption rates and thus the increase in temperature. These specific absorption rates of different parts of exposed rats' brains varied on average by a factor of two and up to a factor of five. This significant variation in heating of different parts of rats' brains is consistent with this model of localized heating in neural circuits. Finally, this model is consistent with a recent article by A. Burlaka, et al., reporting the overproduction of free radicals in quail embryonal cells exposed to cell phone radiation.[9]

This model is similar to the model of Eichler, except that Eichler's model has the induced voltage across the synapse membrane, not along the axon. Eichler estimated that the resistance across a synapse membrane was much greater than the resistance along any other part of the neuron. If this were the case, the induced voltage would go as the radius of the neural circuit squared from Faraday's Law, and the heating would go as the radius to the fourth power, because the number and size of the synapses in a neural circuit would not change with the size of the animal. Given that the axons are long (~1mm) and thin (~1um) and the synapse membranes are thin (~30nm) and wide (~2um), it is not clear to this author whether the resistance at 900 MHz is greater along the axons or across the synapse membranes. Two further comments about the axon vs synapse membrane question. First, the NTP reported the observation of schwannomas—cancers of the neural sheaths rather than cancers located at the synapses; this supports the axon model. Second, given that the male Sprague Dawley rats at age 15 weeks range in weight (and thus size) from 320 grams to 460 grams, it should be possible with sufficient data to distinguish between a square vs a quartic dependence of the incidence of cancer as a function of rat size—assuming heating is the critical factor in carcinogenesis.

This model makes some predictions. Given that the male Sprague Dawley rats are about three times larger than B6C3F1 mice, this model predicts that the cancer rate in male rats is about 9 times greater than in male mice. This model predicts that the cancer rate in male rats should be proportional to the square of the size of the male rats. This model also predicts that the incidence of cancer should increase as the square of the radio wave frequency, assuming that the resistance of the neural circuit does not change with frequency. Hopefully, with greater data from the NTP, these predictions can be tested.

Finally, how are these ideas connected with the relationship between cell phone radiation and cancer in humans? At first glance, one might conclude that the incidence of cancer due to cell phone radiation should be much greater in humans than in rats, because our hearts and brains are so much bigger. Let me caution against such simple logic for the following reasons: 1. Rats are not perfect analogues to humans; there are diseases found in rats that are not found in humans and visa versa. 2. The rats were exposed to radiation 9 hours a day from utero to the time they were killed; the exposure in humans is much less. 3. The radiation actually increased the lives of the rats; if the same is true in humans, many human autopsies would not be looking for these types of cancers.

The question of cancer in humans due to radiofrequency radiation must ultimately be determined by epidemiological studies on humans. Recently, D. Wojcik published a meta-analysis of many different human epidemiological studies and reported that the risk for glioma increased by a factor of 1.9 for cell phone users and mentioned that the French Cerent collaborative case-control study showed the risk for glioma increased by a factor of 2.89 for heavy cell phone users.[10] Very recently, M. Carlberg and L. Hardell published another meta-analysis of many different human epidemiological studies and also reported an increased risk of glioma by a factor of 1.9 for cell phone users.[11] Quoting from their abstract: "RF radiation should be regarded as a human carcinogen causing glioma." These two studies reinforce both the NTP results and the model proposed in this paper.

References

[1]. M. Wyde, M. Cesta, C. Blystone, S. Elmore, P. Foster, M. Hooth, G. Kissling, D. Malarkey, R. Sills, M. Stout, N. Walker, K. Witt, M. Wolfe and J. Bucher, "Report of Partial Findings from the National Toxicology Program Carcinogenesis Studies of Cell Phone Radiofrequency Radiation in Hsd: Sprague Dawley SD Rats (whole body exposures)." http://biorxiv.org/content/early/2016/05/26/055699.full.pdf and https://ecfsapi.fcc.gov/file/10011773529766/EHTrustNTP.pdf.

[2]. R. Melnick, "Cell Phone Radiation Boosts Cancer Rates in Animals; $25 Million NTP Study Finds Brain Tumors," Microwave News, May 25, 2016 http://microwavenews.com/news-center/ntp-cancer-results.

[3] R. K. Adair, "Constraints on Biological Effects of Weak Extremely-Low-Frequency Electromagnetic Fields," Physical Review A 43, 1039 (1992).

[4]. B. J. Feldman, "Physics Déjà Vu," The Physics Teacher 52, 391 (2014).

[5]. D. Eichler, "Nearly Closed Loops in Biological Systems as Electromagnetic Receptors," Bioelectrochemistry and Bioenergetics 42, 227-230 (1997).

[6] F. Barnes and B. Greenebaum, "Some Effects of Weak Magnetic Fields on Biological Systems," IEEE Power Electronics Magazine, March 2016, p. 60.

[7]. D. J. Panagopoulos, O. Johansson and G. L. Carlo, "Polarization: A key Difference between Man-made and Natural Electromagnetic Fields, in regard to Biological Activity," Nature.com/Scientific Reports 5:14914, October 12, 2015.

[8]. C.-K. Chou, A. W. Guy, J. A. McDougall and H. Lai, "Specific Absorption Rate in Rats Exposed to 2,450-MHz Microwaves Under Seven Exposure Conditions," Bioelectromagnetics 6, 73-88 (1985).

[9]. A. A. Burlaka, O. Tsybulin, E. Sidorik, S. Lukin, V. Polishuk, S. Tsehmistrenko and I. Yakymenko, "Overproduction of Free Radical Species in Embryonal Cells Exposed to Low Intensity Radiofrequency Radiation," Experimental Oncology 35, 219-225 (2013).

[10]. D. Wojcik, "Primary Brain Tumors and Mobile Phone Usage," Cancer Epidemiology 44, 123-124 (2016).

[11]. M. Carlberg and L. Hardell, "Evaluation of Mobile Phone and Cordless Phone Use and Glioma Risk Using the Bradford Hill Viewpoints from 1965 on Association and Causation," BioMed Research International 2017 (2017), Article ID 9218486.